# Effective Thermal Conductivity of SrBi$_4$Ti$_4$O$_{15}$-La$_{0.7}$Sr$_{0.3}$MnO$_3$ Oxide composite: Role of Particle Size and Interface Thermal Resistance

Ashutosh Kumar[1,#], Artur Kosonowski[2], Piotr Wyzga[1], Krzysztof T. Wojciechowski[2]

[1]Lukasiewicz Research Network - Cracow Institute of Technology, Zakopianska 73, 30-418 Krakow, Poland
[2]Faculty of Materials Science and Ceramics, AGH University of Science and Technology, 30-059 Krakow, Poland

**Abstract:** We present a novel approach to reduce the thermal conductivity ($k$) in thermoelectric composite materials using acoustic impedance mismatch and the Debye model. Also, the correlation between interface thermal resistance ($R_{int}$) and the particle size of the dispersed phase on the $k$ of the composite is discussed. In particular, the $k$ of an oxide composite which consists of a natural superlattice Aurivillius phase (SrBi$_4$Ti$_4$O$_{15}$) as a matrix and perovskite (La$_{0.7}$Sr$_{0.3}$MnO$_3$) as a dispersed phase is investigated. A significant reduction in the $k$ of composite, even lower than the $k$ of the matrix when the particle size of La$_{0.7}$Sr$_{0.3}$MnO$_3$ is smaller than the Kapitza radius ($a_K$) is observed, depicting that $R_{int}$ dominates for particle size lower than $a_K$ due to increased surface to volume ratio. The obtained results have the potential to provide new directions for engineering composite thermoelectric systems with desired thermal conductivity and promising in the field of energy harvesting.

[#]Email: ashutosh.kumar@kit.lukasiewicz.gov.pl
science.ashutosh@gmail.com

## INTRODUCTION

Thermoelectric (TE) materials enable direct and reversible conversion of heat energy to electrical power; they play an essential role in power generation and solid-state refrigeration [1][2]. The performance of TE materials is quantified using a dimensionless quantity called figure-of-merit (zT) and zT =$S^2$σT/$k$, where S is the Seebeck coefficient, σ is the electrical conductivity, T is the absolute temperature, and $k$ is the thermal conductivity. Oxide materials with excellent heat stability, less-toxic nature, and environment-friendly behavior have received considerable research interest [3][4]. The desired TE parameters like low $k$ (such as in glass), high σ (like in metals), and high S (as in semiconductors) in a single oxide system are challenging to achieve due to the interdependence between S and σ [5] [6]. However, $k$ (= $k_e$ + $k_{ph}$) in oxide materials is dominated by the phonon thermal conductivity ($k_{ph}$), in general, and can be tuned independently as only electronic thermal conductivity ($k_e$) depends on σ according to Wiedemann Franz law. Also, a reduction in $k_{ph}$ is one of the primary requirements to achieve a high zT in TE materials.

There are several processes, including nanostructuring [7], lattice defects [8], mass disorder [9], which have been used to significantly reduce $k_{ph}$. However, phonon scattering *via* lattice defects, impurities, and anharmonicity in a single TE material may also enhance the scattering of electrons that result to lower mobility and electrical conductivity. The composite approach [10] may be promising for the reduction in $k_{ph}$, without reducing the electrical properties, if optimized adequately based on their elastic properties (sound velocity), interface thermal resistance ($R_{int}$), and Kapitza radius ($a_K$) [11].

Based on the available reports in literature, the composite TE materials is an effective strategy to improve the zT *via* reduction of $k_{ph}$ in several systems including $SrTiO_3$-graphene [12], $LaCoO_3$-graphene [13], $CoSb_3$-$WO_3$ [14], $Na_{0.77}CoO_2$-$Ca_3Co_4O_9$ [15], $(Ca_{0.9}Ag_{0.1})_3Co_4O_9$-Ag [16], $Ca_3Co_4O_9$-$La_{0.8}Sr_{0.2}CoO_3$ [17] and also in several other composite systems [18][19]. However, the studies in the literature on the TE composite materials focus on the improvement of transport properties and rarely consider the effect of $R_{int}$ between the phases. However, the role of $R_{int}$ on the $k$ was shown in few composites e.g. SiC-Al [20], diamond-ZnS [11], glass-epoxy [21],

Bi$_2$Se$_3$-PVDF [22] etc. These studies suggest that $R_{int}$ plays a vital role in designing the TE composite materials with desired $k_{ph}$ values promising for TE applications. Also, the correlation between the particle size of the dispersed phase and $R_{int}$ requires additional attention, when designing new TE composites.

Several promising TE properties are demonstrated in oxide systems with a layered structure, like NaCo$_2$O$_4$ [23], Ca$_3$Co$_4$O$_9$ [24], Ruddlesden-Popper phase systems [25]. Recently, an enormous value of S was shown in the Bi$_2$VO$_{5.5}$ phase [26]: an Aurivillius phase (AP) system. AP materials possess a natural superlattice structure [27], having a general formula (Bi$_2$O$_2$)$^{2+}$:(A$_{n-1}$B$_n$O$_{3n+1}$)$^{2-}$, where A is Lanthanides or group II elements, B is transition metals, and n represents the order of BO$_6$ octahedra between (Bi$^2$O$^2$)$^{2+}$ layers. These AP systems possess very high S values due to quantum confinement effect and presence of heavy carriers and low $k$ owing to complex crystal structure [26][28]; however, they possess poor σ and hence less promising for TE applications. These materials can be made suitable for TE applications by adding a conducting phase; however, this may lead to an increase in the $k$ along with σ, if not optimized properly.

In the present study, we show the thermal conductivity ($k$) behavior in a composite that consists of an AP (SrBi$_4$Ti$_4$O$_{15}$) as a matrix and a conducting perovskite (La$_{0.7}$Sr$_{0.3}$MnO$_3$ [29]) system as a dispersed phase over a wide temperature range from 298 K-773 K. In particular, the role of $R_{int}$ along with the different particle sizes of the La$_{0.7}$Sr$_{0.3}$MnO$_3$ on the thermal conductivity of (1-x)SrBi$_4$Ti$_4$O$_{15}$ - (x)La$_{0.7}$Sr$_{0.3}$MnO$_3$ (0≤x≤1.00) composite is discussed. The experimental results obtained for the composites have been compared with Bruggeman's model within effective media theory.

## EXPERIMENTAL SECTION

The polycrystalline SrBi$_4$Ti$_4$O$_{15}$ (SBTO) and La$_{0.7}$Sr$_{0.3}$MnO$_3$ (LSMO) were separately synthesized using a standard solid-state route [30][31]. For the synthesis of SBTO, the stoichiometric amount of SrCO$_3$, Bi$_2$O$_3$, and TiO$_2$ were mixed using a mortar-pestle in a liquid medium. The mixed powders were calcined at 1093 K and 1173 K for 12 hours with intermediate grinding with 3 K/min heating and cooling rate. The LSMO sample was prepared by mixing the stoichiometric amount of $La_2O_3$, $SrCO_3$, and $Mn_2O_3$ following the process, as mentioned above. The mixed LSMO precursors were calcined at 1473 K for 12 hours with slow heating and cooling rate (3 K/min). Both the calcined powders SBTO and LSMO were ground individually to make homogeneous powder. Next, the Aurivillius phase-perovskite composites (1-x)SBTO - (x)LSMO were prepared by mixing the synthesized LSMO powder with SBTO (0.00 ≤x≤1.00), x represents the variation in the weight for each phase in the composite. The particle size for SBTO and LSMO is similar (~2-3 µm) for this series. Further, LSMO powders were ball-milled in a zirconia jar using a planetary ball mill with 320 rpm for 10 hours and 24 hours to achieve two different particle sizes of LSMO (~500 nm and ~100 nm). The ball (zirconia) to powder weight ratio was maintained to 10:1. Two composite samples having the same weight composition (0.50)SBTO - (0.50)LSMO with ~500 nm, and ~100 nm particle sizes of LSMO were synthesized. All the mixtures were pressed in the form of pellets having a 10 mm diameter and 1.5-2 mm thickness by applying a 40MPa pressure for 60 sec. The composite pellets were sintered using the conventional sintering process at 1123 K for 6 hours with a 3K/min cooling and heating rate. The structural characterization of the composite was done using the x-ray diffraction (XRD) technique (Cu-K$_\alpha$, λ =1.5406 Å), followed by two-phase Rietveld refinement using Fullprof$^{TM}$ software [32]. The surface of the sample for microstructure investigation was polished using an automatic grinding/polishing machine. The surface morphology of the as-sintered and polished sample was observed using a scanning electron microscope (SEM, NOVA NANO 200, FEI EUROPE Company). The thermal conductivity ($k = Dc_P\rho$) for all the samples were measured in a temperature range from 298 K-773 K. The thermal diffusivity ($D$) was measured using laser flash analysis (LFA) in the Ar atmosphere. The specific heat capacity ($c_P$) was calculated using Dulong-Petit law. The bulk density ($\rho$) was calculated using sample mass and its geometrical volume. The error in thermal conductivity measurement is ~7%. Sound velocity was measured using ultrasonic flaw detector EPOCH 3 (Panametrix) at room temperature.

## RESULTS AND DISCUSSION

### A. Structural Characterization

Figure 1 represents the X-ray diffraction pattern for the sintered (1-x)SBTO - (x)LSMO composite. The Bragg's position and Miller indices for SBTO and LSMO are also marked. The existence of both the phases in the composite is shown in Fig. 1. No impurity phases in the diffraction pattern are observed within the sensitivity of the XRD technique. The characteristic 2θ peak at 30.19° [119] and doublet peaks at 32.64° (110), 32.84° (104) corresponds to SBTO and LSMO, respectively.

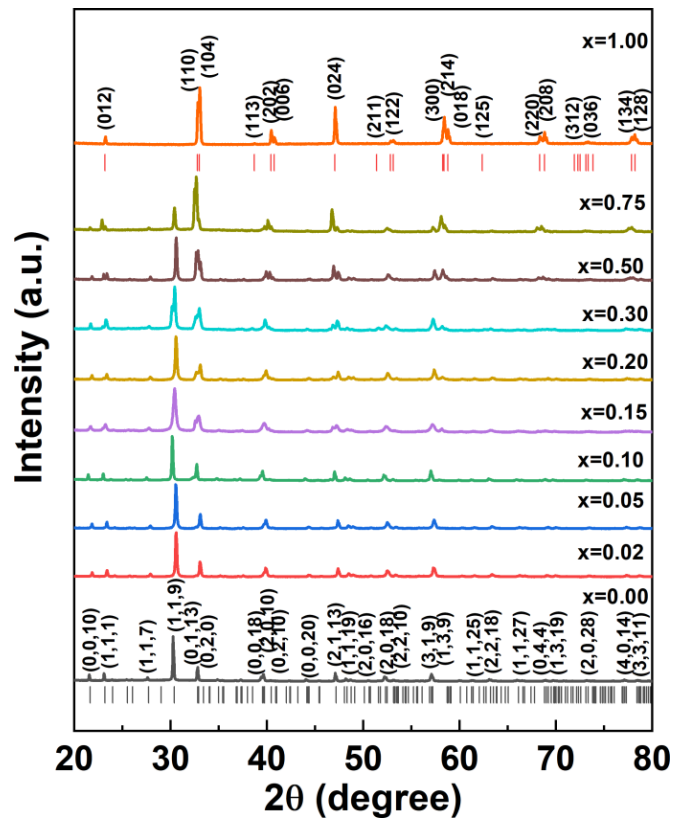

FIG. 1: X-ray diffraction pattern for (1-x)SBTO - (x)LSMO composite. The Bragg's position and Miller indices for both the phases are marked.

The diffraction intensity due to LSMO increases with an increase in LSMO weight fraction in the composite. The diffraction pattern for SBTO and LSMO is in line with the ICDD file no. 043-0973 and 070-8668 respectively [33][34]. The quantitative analysis of the x-ray diffraction pattern is done via Rietveld refinement using Fullprof$^{TM}$ software. The single-phase Rietveld refinement for SBTO and LSMO is performed using tetragonal (A21am) and rhombohedral (R-3c) structure, respectively.

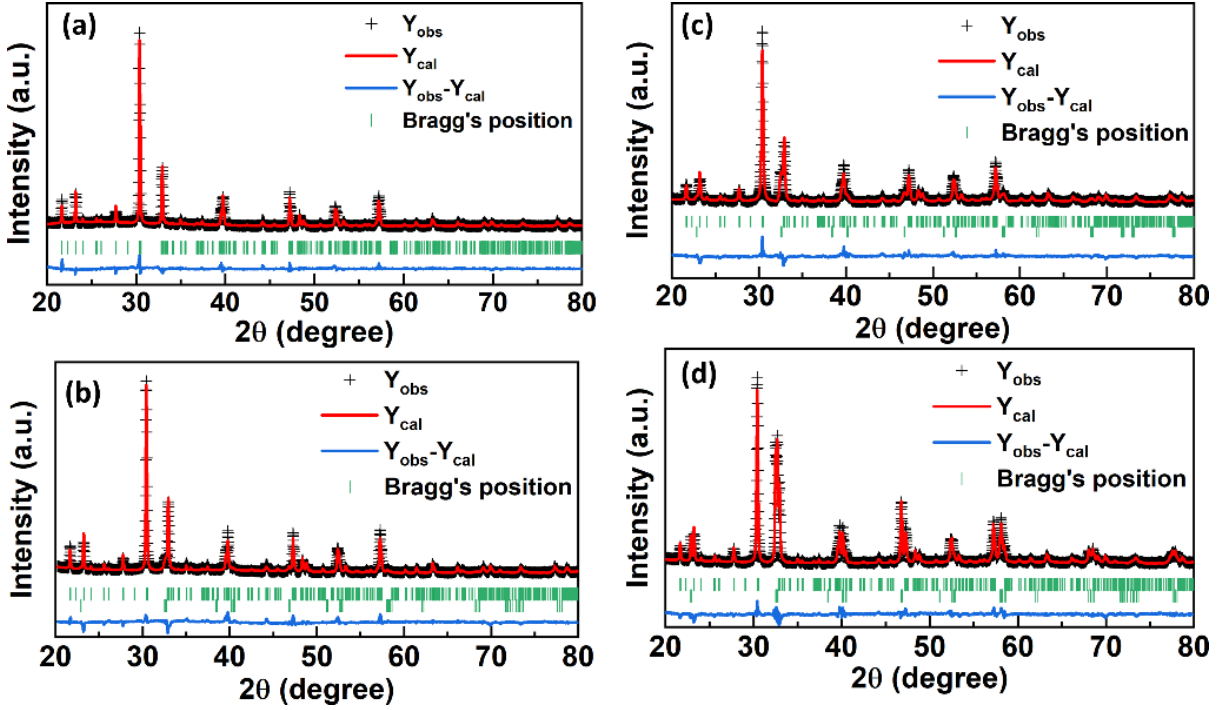

FIG. 2: Rietveld refinement pattern for (1-x)SBTO - (x)LSMO composite for (a) x=0.00 (b) 0.10 (c) 0.20 and (d) x=0.50.

TABLE I: Rietveld refinement parameters ($R_{exp}$, $R_{wp}$, and $\chi^2$) and weight fraction obtained for the (1-x)SBTO - (x)LSMO composite. The lattice parameters for LSMO is a=5.462 Å, c=13.147 Å and SBTO is a=5.447 Å, c=41.165 Å.

| Sample composition | $R_{exp}$ | $R_{wp}$ | $\chi^2$ | weight fraction | |
|---|---|---|---|---|---|
| | | | | SBTO (%) | LSMO (%) |
| x=0.00 | 5.33 | 7.35 | 1.90 | 100 | 0 |
| x=0.02 | 5.46 | 7.39 | 1.83 | 98.2 | 1.8 |
| x=0.05 | 5.40 | 7.45 | 1.90 | 95.9 | 4.1 |
| x=0.10 | 5.82 | 7.65 | 1.73 | 90.2 | 9.8 |
| x=0.15 | 5.65 | 7.70 | 1.86 | 86.3 | 13.7 |
| x=0.20 | 6.88 | 8.59 | 1.56 | 81.2 | 18.8 |
| x=0.30 | 6.02 | 8.24 | 1.87 | 71.0 | 29.0 |
| x=0.50 | 6.93 | 9.11 | 1.73 | 50.4 | 49.6 |
| x=1.00 | 5.69 | 7.12 | 1.57 | 0 | 100 |

Further, the obtained lattice parameters for SBTO and LSMO have been used to perform two-phase Rietveld refinement. The refinement parameters obtained for the composite samples are shown in Table I. The weight fraction obtained from the Rietveld refinement is in good agreement with the nominal composition of both the phases in the composite. The atomic positions and their occupancies obtained from the refinement for Sr, Bi, Ti, and O in SBTO and La, Sr, Mn, and O in LSMO are shown in Table II.

The bulk density of the sample is measured using sample mass and its geometrical volume. The bulk, theoretical, and relative density along with porosity for all the samples are shown in Table III. The theoretical densities have been calculated via the volume-weighted arithmetic mean. The relative density for all the samples lies in the range of 73%-83%.

TABLE II: Atomic positions and occupancies of different elements in SBTO and LSMO used for Rietveld refinement of the XRD pattern for the SBTO-LSMO composite sample.

| SBTO (A21am) | | | | |
|---|---|---|---|---|
| Atom | X | Y | Z | Occupancy |
| Sr(1) | 0.417(9) | 0.270(9) | 0.000(0) | 0.091 |
| Sr(2) | 0.000(0) | 0.000(0) | 0.000(0) | 0.029 |
| Bi(1) | 0.417(9) | 0.270(0) | 0.000(0) | 0.391 |
| Bi(2) | 0.444(1) | 0.268(0) | 0.105(6) | 0.832 |
| Bi(3) | 0.435(3) | 0.239(8) | 0.223(0) | 0.975 |
| Ti(1) | 0.383(4) | 0.201(4) | 0.440(8) | 0.998 |
| O(1) | 0.316(4) | 0.407(5) | 0.500(0) | 0.661 |
| O(2) | 0.504(2) | 0.418(2) | 0.046(2) | 0.798 |
| O(3) | 0.163(8) | 0.011(0) | 0.435(2) | 1.592 |
| O(4) | 0.737(0) | 0.842(0) | 0.155(1) | 1.165 |
| O(5) | 0.308(1) | 0.200(0) | 0.322(0) | 1.244 |
| O(6) | 0.648(7) | 0.539(3) | 0.244(9) | 1.265 |
| O(7) | 0.064(0) | 0.277(9) | 0.034(5) | 1.655 |
| O(8) | 0.079(3) | 0.155(1) | 0.167(3) | 1.135 |
| LSMO (R-3c) | | | | |
| Atom | X | Y | Z | Occupancy |
| La | 0.000(0) | 0.000(0) | 0.250(0) | 0.112 |
| Sr | 0.000(0) | 0.000(0) | 0.250(0) | 0.052 |
| Mn | 0.000(0) | 0.000(0) | 0.000(0) | 0.180 |
| O | 0.447(8) | 0.000(0) | 0.250(0) | 0.557 |

Fig. 3(a-l) show the scanning electron microscope image for as-sintered (1-x)SBTO - (x)LSMO composite and polished surfaces. The particle sizes for LSMO lies in the range of 1-3 µm and having a spherical shape; however, SBTO samples are lobe-shaped. The ball-milled LSMO samples are in the range of 450-600 nm and 80-135 nm for 10 hours and 24 hours of ball-milling, respectively [35]. SEM images of the polished pellets of composite samples show the presence of both kinds of phases.

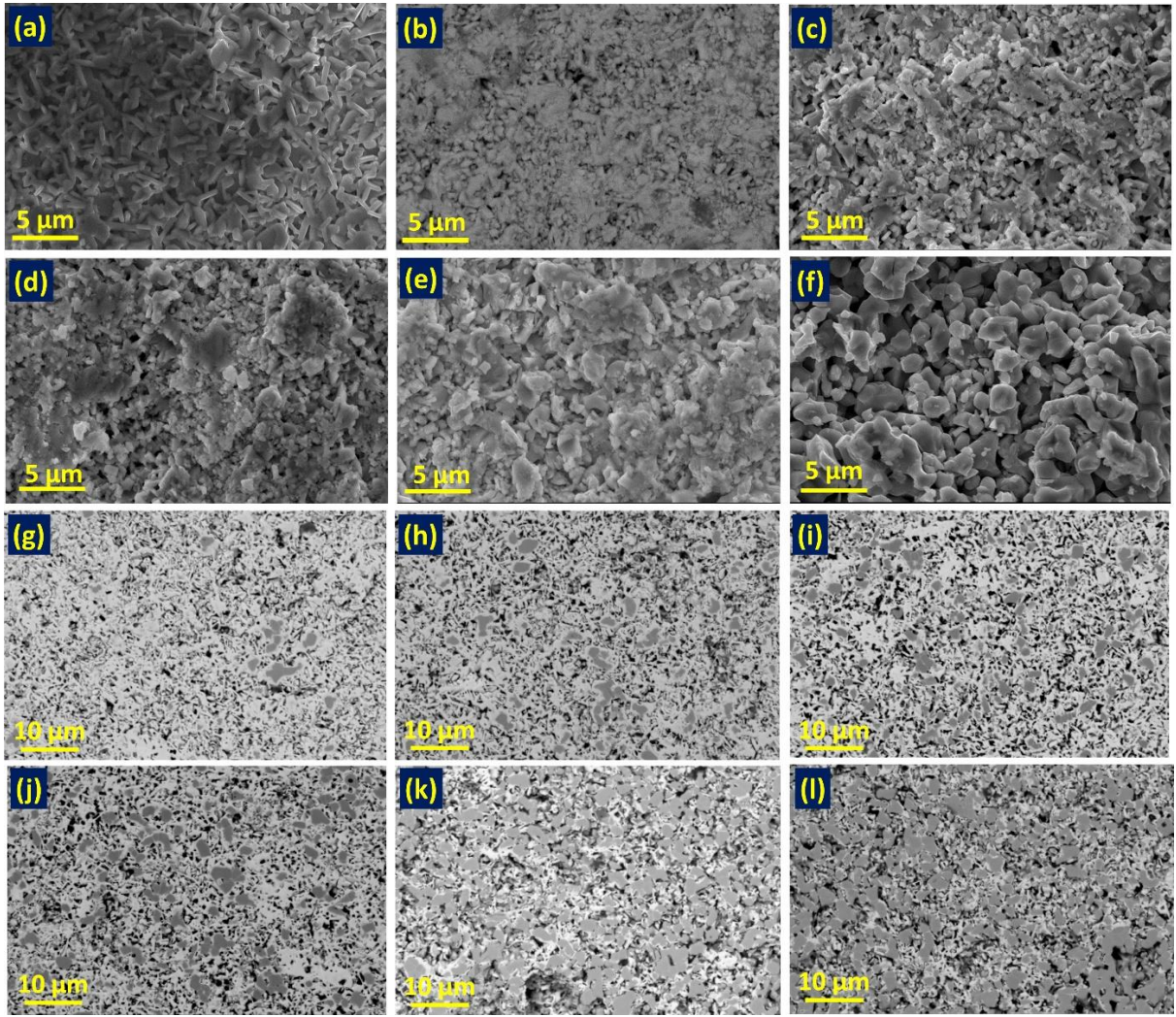

FIG. 3: Scanning electron microscope (SEM) image for as sintered (1-x)SBTO - (x)LSMO composite pellets for (a) x=0.00 (b) 0.05 (c) 0.10 (d) x=0.20 (e) x=0.50 and (f) x=1.00. The SEM images for the (1-x)SBTO - (x)LSMO composite pellets after polishing for (g) x=0.05, (h) x=0.10, (i) x=0.15, (j) x=0.20, (k) x=0.50, and (l) x=0.75.

TABLE III: Bulk density (ρ), theoretical density ($\rho_{th}$), relative density ($\rho_{rel}$), and porosity for (1-x)SBTO - (x)LSMO composite.

| Sample | ρ (g/cm³) | $\rho_{th}$ (g/cm³) | $\rho_{rel}$ (%) | Porosity (%) |
|---|---|---|---|---|
| x=0.00 | 5.65 | 7.44 | 76 | 24 |
| x=0.02 | 5.55 | 7.42 | 75 | 25 |
| x=0.05 | 5.47 | 7.39 | 74 | 26 |
| x=0.10 | 5.48 | 7.33 | 75 | 25 |
| x=0.15 | 5.53 | 7.28 | 76 | 24 |
| x=0.20 | 5.69 | 7.23 | 79 | 21 |
| x=0.30 | 5.49 | 7.13 | 77 | 23 |
| x=0.50 | 5.07 | 6.94 | 73 | 27 |
| x=0.75 | 4.99 | 6.71 | 74 | 26 |
| x=1.00 | 5.38 | 6.50 | 83 | 17 |

## B. Thermal Properties

Figure 4(a) shows the thermal conductivity ($k$) behavior of the (1-x)SBTO - (x)LSMO (0.00 ≤ x ≤ 1.00) composite as a function of temperature. A low value of $k$ (∼ 0.5 W/mK) is observed at 298 K for the SBTO sample. It is due to the presence of $(Bi_2O_2)^{2+}$ interslab layer between $(SrBi_2Ti_4O_{13})^{2-}$ layer in the natural superlattice Aurivillius phase system [26]. The $k$ increases with the increase in the LSMO weight fraction.

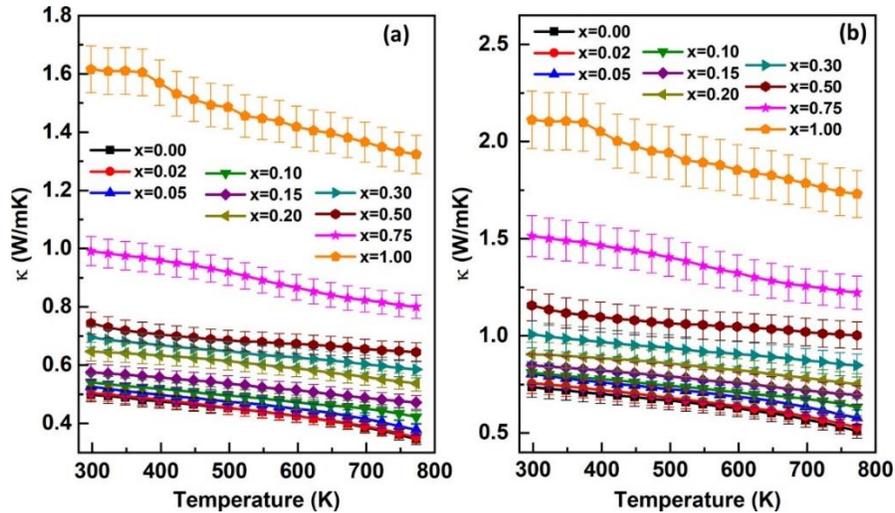

FIG. 4: Thermal conductivity ($k$) as a function of temperature for the (1-x)SBTO - (x)LSMO composite (a) without porosity correction, and (b) with porosity correction.

As shown in Table III, the relative densities for all the samples are in the range of 73-83%. Thus, there is an influence of porosity on $k$ in the composite samples. To avoid the contribution of porosity on $k$ in the composite samples, we performed porosity correction to $k$ of the composites using the Maxwell–Eucken model [36]:

$$\frac{k_d}{k} = \left(\frac{1-P}{1+(\beta-1)P}\right) \quad (1)$$

where $k$ is effective thermal conductivity in a porous medium, $k_d$ is the thermal conductivity of the material with 100% density, P is the volume fraction of pores, β is a constant and depends on the shape and size of the pores called a property-pore-shape coefficient (β=1.5 for spherical shape [37]). Figure 4(b) shows the thermal conductivity with porosity correction as a function of temperature for (1-x)SBTO - (x)LSMO composite. The $k$ for pure SBTO sample (after porosity correction) is ∼ 0.73 W/mK at 298 K. The electrical conductivity in the pure SBTO is small [37]. It indicates that the corresponding electronic contribution to the thermal conductivity ($k_e$) is negligible. Hence, the phonon thermal conductivity ($k_{ph}$) dominates the total thermal conductivity. The value of $k$ decreases with the increase in temperature for all the samples. This decrease in $k$ suggests that $k_{ph}$ decreases with the increase in temperature and can be attributed to the improved phonon scattering due to a decrease in the mean free path of phonons at higher temperatures [38]. Further, $k$ increases with the increase in the LSMO weight fraction in the composite. The $k_e$ due to LSMO is also not significant, and hence the $k$ in the composite is also dominated by $k_{ph}$. Further, the monotonous increase in the $k$ of the composite as a function of LSMO volume fraction (f) is explained using the $R_{int}$ between these two phases of the composite. It is worth noting that the LSMO weight fraction (x) has been converted to its corresponding volume fraction (f) for the detailed analysis using the Bruggeman model. The $R_{int}$ is calculated using acoustic impedance mismatch (AIM) and the Debye model [39] [40]. Further, $k$ as a function of LSMO volume fraction (f) is fitted using Bruggeman's model [41].

TABLE IV: Measured values of bulk density (ρ), sound velocity (𝑣) used in acoustic impedance model (AIM) to calculate the acoustic impedance (Z) and Transmission coefficient (p) of phonons for SBTO and LSMO samples.

| Sample Name | ρ (g/cm³) | $v$ (m/sec) | | Z (kg/(m²s)) | | p (%) |
|---|---|---|---|---|---|---|
| | | $v_t$ (Trans.) | $v_t$ (Long.) | Trans. | Long. | |
| SBTO | 5.65 | 1600 | 2610 | 14747 | 9040 | 99.2 |
| LSMO | 5.38 | 2110 | 3170 | 17055 | 11352 | |

## 1. Acoustic Impedance Mismatch (AIM) Model

The acoustic impedance mismatch (AIM) model provides a basis for understanding the interface thermal resistance ($R_{int}$) between the phases of a composite. Also, AIM can be used to calculate the probability of phonons transmission at the interface between the two phases. Let us consider the phonons in a polycrystalline composite system to travel through one material to another and confront the interface between the two materials. The fraction of phonons (q) having the incident angle below a critical value can be calculated using the following equation [11]:

$$q = \frac{1}{2}\left(\frac{v_m}{v_d}\right)^2 \quad (2)$$

where $v_m$ and $v_d$ are the sound velocities of the matrix (SBTO) and the dispersed phase (LSMO), respectively. The sound velocities measurement of these samples results in a $q$ value of 0.288. It indicates that 28% of these phonons have the incident angle within the critical angle. Further, the transmission coefficient (p) of the phonons at the interface of two materials can be estimated using the following equation [42]:

$$p = \frac{4Z_A Z_B}{(Z_A + Z_B)^2} \quad (3)$$

where $Z_i = v_i \rho_i$ is acoustic impedance, $\rho_i$ is the density, and $v_i$ is the velocity of the i$^{th}$ material. The measured $v$ and $\rho$ are used to calculate the acoustic impedance for both the materials. The acoustic impedance for both the phases gives a $p$ value of 0.9921. It shows that the phonons which incident on the interface below the critical angle, 99.2% of these can cross the interface. Hence, the transmission probability ($\eta = pq$) for these SBTO- LSMO interface is 0.2852. Further, the $R_{int}$ for the SBTO-LSMO composite is calculated using the Debye model given by the equation:

$$R_{int}^D = \frac{4}{\rho c_p v_D \eta} \tag{4}$$

where $c_p$ is the specific heat capacity, $v_D \left( = \left( \frac{3}{\frac{1}{v_l^3} + \frac{2}{v_t^3}} \right)^{1/3} \right)$ is the Debye velocity, $v_l$ and $v_t$ are the sound velocities in a longitudinal and transverse direction, respectively. Using the values of $c_p$, ρ, $v_D$, and η, the interface thermal resistance ($R_{int}^D$) estimated using the Debye model is 4.77 × 10$^{-7}$ m²K/W. Further, a theoretical estimation of Kapitza radius ($a_K = R_{int}^D k_m$) can be made, where $k_m$ is the thermal conductivity of the matrix (SBTO, $k_m$=0.735 W/mK) and $R_{int}^D$=4.77× 10$^{-7}$ m²K/W at 298 K. It gives a Kapitza radius of ∼ 350 nm at 298 K.

**2. Bruggeman's Model:**

Further, for additional validation of the experimental results of $k$, we used Bruggeman's symmetrical model that considers the individual phases in the composite fill the media and are in contact with each other. This model predicts the thermal conductivity of the composite when the thermal conductivity and volume fraction of individual components are known. Bruggeman's symmetrical model is given by [43]

$$\frac{f_A(k_A-k)}{k_A+(d-1)} + \frac{f_B(k_B-k)}{k_B+(d-1)} = 0 \tag{5}$$

where $k_A$, $k_B$, and $k$ are thermal conductivity of material 1, material 2, and composite respectively; $f_A$, $f_B$ are the volume fractions of material 1 and 2, respectively, and $d$ is the dimensionality of the considered system. Assuming $d$=3 and solving Eq. (5) for $k$ one can obtain the formula for thermal conductivity of composite:

$$k = \frac{2f_A k_A + 2f_B k_B - f_B k_A - f_A k_B + \sqrt{(f_A(k_B-2k_A)+f_B(k_A-2k_B))^2 + 8k_A k_B}}{4} \tag{6}$$

The calculated $k$ (solid lines) as a function of LSMO volume fraction (f) considering Bruggeman's symmetrical model along with experimental values (symbols) at different temperatures is shown in Fig. 5(a).

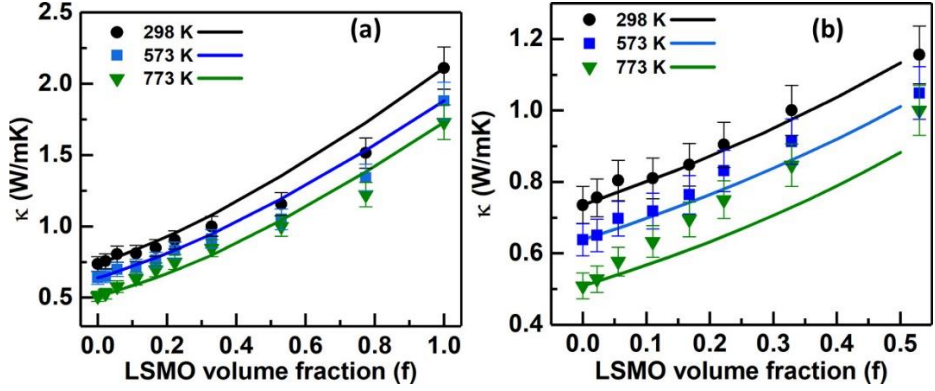

FIG. 5: Thermal conductivity ($k$) of the (1-x)SBTO - (x)LSMO composite as a function of LSMO volume fraction (f) for different temperatures. The weight fraction (x) has been changed to volume fraction (f) using the density of each phase. The solid lines represent the $k$ values according to (a) Bruggeman's symmetrical model, and (b) Bruggeman's asymmetrical model at each temperature. Symbols represent the experimental values of $k$.

Next, the effect of $R_{int}$ and $a_K$ on $k$ of the composite is discussed. Let us consider Bruggeman's asymmetrical model that considers the presence of $R_{int}$ between the two phases in the composite. This model assumes that one material in a composite is a continuous matrix, and the second phase in the composite consists of spherical particles dispersed uniformly within the matrix. Bruggeman's asymmetrical model is given by the formula [43]

$$(1-f)^3 = \left(\frac{k_m}{k}\right)^{\frac{1+2\alpha}{1-\alpha}} \left(\frac{k-k_d(1-\alpha)}{k_m-k_d(1-\alpha)}\right)^{\frac{3}{1-\alpha}} \quad (7)$$

where $k_d$ and $k$ are the thermal conductivity of the dispersed phase (LSMO) and the composite, respectively. The $f$ is the volume fraction of dispersed phase and $\alpha = a/a_K$, where $a$ is the particle/inclusion size of the dispersed phase. Owing to the asymmetry, this model is valid up to a 50% volume fraction of the dispersed phase (i.e., $f \leq 0.50$.). Figure 5(b) represents the $k$ as a function of LSMO volume fraction ($f$) in the (1-x)SBTO - (x)LSMO composite for different temperatures following the Bruggeman's asymmetrical model (solid lines) along with experimental $k$ (symbols) at each temperature. The statistical parameter obtained from the fitting of $k$ using Bruggeman's symmetrical and asymmetrical model is shown in Table V. Bruggeman's symmetrical model fits better at higher temperatures (i.e., 773 K). Bruggeman's asymmetrical model fits well at 298 K and deviates at a higher temperature. It can be attributed to the constant value of $R_{int}$ and $a_K$ used for the calculation of $k$. An increase in $k$ of the composite with the increase in the LSMO volume fraction (f) is observed, and this suggests that the $R_{int}$ is not significant to reduce $k$ when the particle size of LSMO is larger than $a_K$.

TABLE V: The statistical parameters, coefficient of determination ($R^2$), and mean square weighted deviation ($\chi^2$), obtained from fitting of experimental $k$ using two different Bruggeman's models.

| Bruggeman's Model | 298K | | 573K | | 773K | |
|---|---|---|---|---|---|---|
| | $R^2$ | $\chi^2$ | $R^2$ | $\chi^2$ | $R^2$ | $\chi^2$ |
| Symmetric | 0.907 | 1.804 | 0.945 | 1.137 | 0.961 | 1.064 |
| Asymmetric | 0.979 | 0.118 | 0.945 | 0.413 | 0.783 | 3.011 |

Figure 6 shows the $k$ as a function of temperature for (1-x)SBTO - (x)LSMO composite for x=0.50 with a different particle size distribution of LSMO (3 µm, 500 nm, and 100 nm). The $k$ of the composite reduces significantly when the particle size of LSMO is lowered from 3 µm to 500 nm to 100 nm. The corresponding $k$ decreases from 1.15 W/mK (3 µm) to 0.71 W/mK (500 nm) to 0.39 W/mK (100 nm) at 298 K. The $k$ for the composite with particle size ~100 nm is even lower than the $k$ of the matrix (0.73 W/mK). This significant reduction in the $k$ may be due to the prominent effect of $R_{int}$ when the particle size of LSMO reduces. The reduction in particle size of LSMO increases the surface to volume ratio and hence the $R_{int}$ becomes effective to lower the $k$.

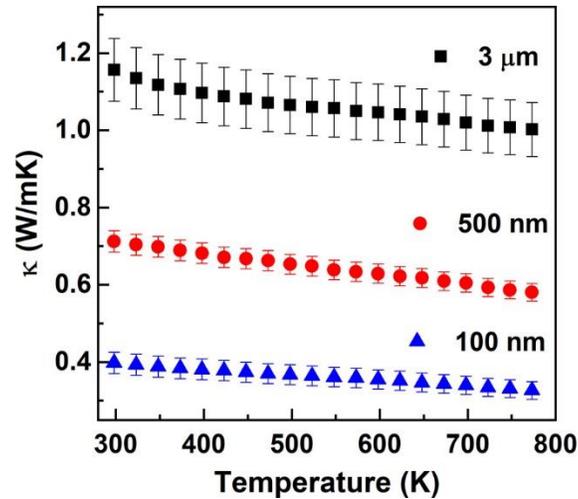

FIG. 6: Thermal conductivity ($k$) as a function of temperature for (1-x)SBTO - (x)LSMO composite for x=0.50 with a different particle size of LSMO.

Figure 7(a) represents the schematic of an acoustic impedance mismatch model. The probability of phonon transmission and phonon reflection is shown as a function of mismatch in acoustic impedance of two materials. The composite materials may have a lower $k$ when the acoustic impedance mismatch ($Z_A/Z_B$) is large, as shown in Fig. 7(a), due to an increase in the phonon reflection coefficient (1-$p$). It is noted that $R_{int}$ and $a_K$ are correlated, a high $R_{int}$ can reduce the $k$ of a composite for large particle size (1-2 µm) of the dispersed phase [11]. However, for smaller $R_{int}$ it is also possible to reduce the $k$ with a smaller particle size of the dispersed phase.

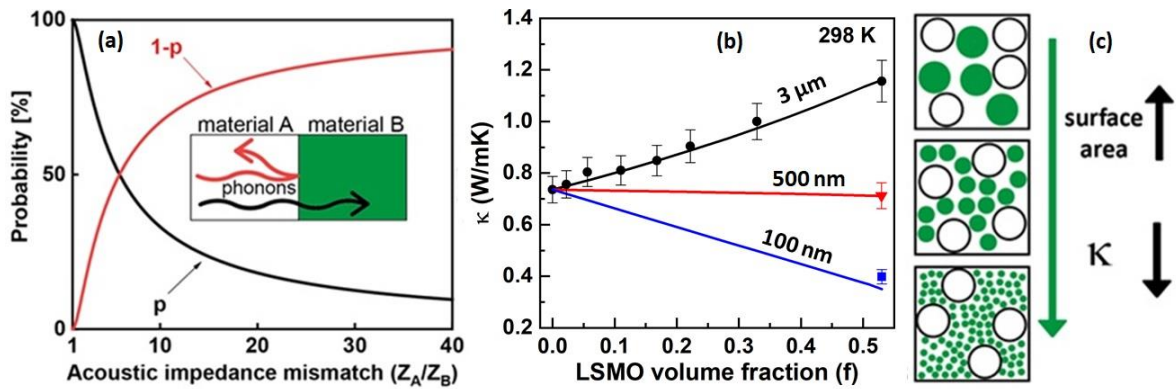

FIG. 7: (a) Schematic of acoustic impedance mismatch (AIM) model, represents the probability of phonon transmission ($p$) and reflection (1-$p$) at the boundary of two materials having a different acoustic impedance. (b) Thermal conductivity ($k$) of the composite as a function LSMO volume fraction (f) with different particle sizes at 298 K. Solid lines represent the $k$ according to Bruggeman's asymmetrical model, and symbols represent the experimental value. (c) Representation of composite samples with a different particle size of LSMO (filled circles) is shown.

Figure 7(b) shows the $k$ calculated for a different particle/inclusion size of LSMO using Bruggeman's asymmetrical model (solid lines). As shown in Fig. 7(b), when the particle size of LSMO reaches the threshold value of $a_K$ (~500 nm), the $k$ remains almost constant and decreases when the particle size is smaller than $a_K$. This model predicts that $k$ of the composite can be lower than $k_m$ when the particle size of the dispersed phase is smaller than $a_K$ [44]. The experimentally observed $k$ (shown by symbols with respective error bars) for the composite with a different particle size of LSMO at 298 K fits the calculated values. A decrease in the $k$ of the composite, lower than $k_m$, is obtained when the particle size of LSMO is smaller than $a_K$. This decrease in $k$ is understood as following: as the particle size of the dispersed phase decreases to a threshold value ($a_K$), the contact area between phases increases due to an increase in surface to volume ratio (schematic is shown in Fig.7(c)), and $R_{int}$ between the phases becomes more prominent. When the particle size of the LSMO is smaller than $a_K$, even the small value of $R_{int}$ is sufficient to reduce the $k$ of the composite.

Figure 8(a) depicts the calculated value of $k$ from Bruggeman's asymmetrical model for (1-x)SBTO - (x)LSMO composite with different x values at 298 K. The $k$ for the composite decreases with the reduction in the particle/inclusion size of LSMO, due to a larger contact area between the matrix and dispersed phase. We obtained a specific particle size at which the $k$ of the composite is lower than the $k$ of the matrix, and this is called $a_K$. Also, $a_K$ is constant for all the composites, as it only depends on the $R_{int}$ and $k_m$. The $a_K$ for the SBTO-LSMO composite is ~520 nm, which is higher than the value calculated from the Debye model, as shown earlier. A similar difference in the theoretical and experimental value of $a_K$ is also shown by Every et al., in ZnS/diamond composite [11]. The difference in $a_K$ obtained from the Debye model and the Bruggeman asymmetrical model may be due to their different assumptions. The Debye model ignores the effect of sound velocity dispersion and hence underestimates the $R_{int}$ and hence $a_K$ [11].

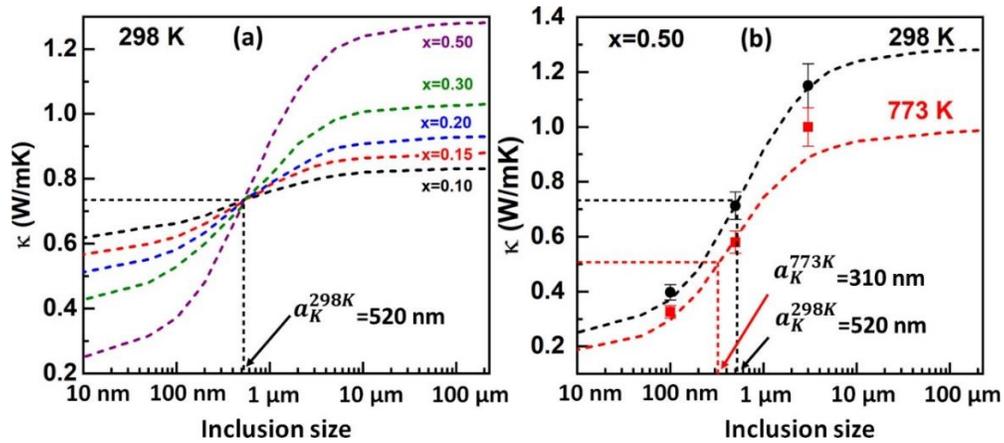

FIG. 8: (a) Calculated thermal conductivity ($k$) of (1-x)SBTO - (x)LSMO composites obtained from Bruggeman's asymmetrical model as a function of the LSMO inclusion size at 298 K. (b) Calculated thermal conductivity ($k$) for 298 K and 773 K for x=0.50. Symbols represent the experimental $k$ value for different LSMO particle size. The Kapitza radius for (1-x)SBTO - (x)LSMO composite for different temperatures is marked.

Figure 8(b) shows the $k$ of the (1-x)SBTO - (x)LSMO composite with x=0.50 as a function of the particle size of LSMO at two different temperatures (dashed lines) calculated from the Bruggeman's asymmetrical model and the symbols represent the experimental value. The experimental value of $k$ for different particle/inclusion sizes of LSMO is in agreement with the Bruggeman's asymmetrical model. The $k$ at 773 K is deviated from the calculated value and is because the $R_{int}$ may vary with the increase in temperature. The $a_K$ decreases at a higher temperature, and this may be due to the decrease in the mean free path of the phonons at high temperatures, which reduces $k_m$ [45]. The results obtained in the present study have the potential to engineer new thermoelectric composite with reduced thermal conductivity values, as desired for promising TE materials.

## CONCLUSION

The role of interface thermal resistance and particle size on the thermal conductivity of (1-x)SrBi$_4$Ti$_4$O$_{15}$ - (x)La$_{0.7}$Sr$_{0.3}$MnO$_3$ (0≤x≤1.00) composite is investigated. The x-ray diffraction (XRD) analysis reveals the presence of individual phases of SrBi$_4$Ti$_4$O$_{15}$ and La$_{0.7}$Sr$_{0.3}$MnO$_3$ (LSMO) in the composite. The two-phase Rietveld refinement further confirms the nominal composition of the individual phases of the composite. The thermal conductivity of the composite increases monotonously with the increase in the LSMO weight fraction (x) and is explained using the acoustic impedance mismatch model. The $k$ as a function of LSMO volume fraction (f) follows Bruggeman's symmetrical model that neglects the presence of $R_{int}$ when the particle size of LSMO is larger than $a_K$ (~520 nm) indicating the inefficiency of $R_{int}$. Further the concept of Kapitza radius ($a_K$) and interface thermal resistance ($R_{int}$) between the phases is realized, and a significant reduction in $k$ of the composite is obtained when the particle size of LSMO is smaller than $a_K$, which is further supported by Bruggeman's asymmetrical model. It indicates that even a small $R_{int}$ can significantly reduce the $k$ of the composite when the particle size of LSMO is smaller than $a_K$. The $k$ for the composite as a function of different particle/inclusion sizes of LSMO shows that $a_K$ is constant for all the samples as it only depends on the $R_{int}$ and $k$ of the matrix. However, $a_K$ decreases with increasing temperature and is ascribed to the reduction of the mean free path of phonons at higher temperatures. This study suggests that the thermal conductivity of a composite can be significantly reduced by enhanced $R_{int}$ between the phases along with the reduced particle size of the dispersed phase. Also, this study paves a new direction to engineering thermoelectric composite materials with reduced thermal conductivity useful to promising thermoelectric applications.

## ACKNOWLEDGMENTS

This research was financed by the 'New approach for the development of efficient materials for direct conversion of heat into electricity' TEAM-TECH/2016-2/14 project, which is carried out within the TEAM-TECH program of the Foundation for Polish Science co-financed by the European Union under the European Regional Development Fund. The beneficiary of this project is The Lukasiewicz Research Network - Cracow Institute of Technology in Krakow (Poland).